\documentclass{article}
\usepackage{amsfonts,amssymb, amsmath,mathrsfs,bigints}
\usepackage[english]{babel}
\usepackage{wrapfig}
\usepackage{graphicx}

\def\nn{\nonumber }
\def\bq{ \begin{equation} }
\def\eq{ \end{equation} }
\def\ben{ \begin{eqnarray} }
\def\en{ \end{eqnarray} }

\def\g{{\gamma}}

\newtheorem{prop}{Proposition}

\newtheorem{re}{Remark}

\begin{document}


\title{On the nonholonomic St\"{u}bler model}
\author{A. V. Tsiganov\\
\it\small
St.Petersburg State University, St.Petersburg, Russia\\
\it\small andrey.tsiganov@gmail.com}

\date{}
\maketitle

\begin{abstract}
We discuss two constructions of the Poisson bivectors based on the Euler-Jacobi theorem  for the nonholonomic St\"{u}bler model, which describes rolling  without sliding of a uniform ball on a cylindrical surface.
 \end{abstract}

\section{Introduction}
\setcounter{equation}{0}
Let us consider the following system of differential equations
\[
\dfrac{\mathrm dx_1}{X_1}=\dfrac{\mathrm dx_2}{X_2}=\cdots=\dfrac{\mathrm dx_k}{X_k}\,,
\]
 where $X_i$ are the functions of variables $x_1,\ldots,x_k$. If we know the Jacobi multiplier $\mu $ defined by
\begin{equation}\label{y-mera}
\dfrac{\partial \mu X_1}{\partial x_1}+\dfrac{\partial \mu X_2}{\partial x_2}+\cdots+\dfrac{\partial \mu X_k}{\partial x_k}=0\,,
\end{equation}
and know $k-2$ independent first integrals $H_1,\ldots, H_{k-2}$, we can integrate this system by quadratures.

Namely, according to the Euler-Jacobi theorem, in this case we can introduce new variables $y_1,\ldots,y_k$ by rule
 \[
 y_1=x_1,\qquad y_2=x_2,\qquad y_3=H_1,\quad\ldots\quad,\, y_k=H_{k-2}\,.
 \]
In $y$-variables the initial system of equations has the following form
\begin{equation}\label{eqy-in}
\dfrac{\mathrm dy_1}{Y_1}=\dfrac{\mathrm dy_2}{Y_2}=\dfrac{\mathrm d y_3}{Y_3}=\cdots=\dfrac{\mathrm dy_k}{Y_k}\,,\qquad\mbox{where}\qquad Y_i=0\,,\quad i\geq 3\,,
\end{equation}
or
\[Y_2\mathrm dy_1-Y_1\mathrm dy_2=0\,.\]
Using the definition (\ref{y-mera}) of the Jacobi multiplier $\mu$
\[
\dfrac{\partial \mu Y_1}{\partial y_1}+\dfrac{\partial \mu Y_2}{\partial y_2}=0\,,
\]
we can prove that $\mu (Y_2\mathrm dy_1-Y_1\mathrm d y_2)$ is the total differential. So, there is one more independent first integral
\[
H_{k-1}=\int \mu (Y_2\mathrm dy_1-Y_1\mathrm d y_2)\,.
\]
Of course, this only a formal definition of the integral, which may be not well defined on the whole phase space.

Nevertheless, it is well-known that any $k$-dimensional dynamical system with the functionally in\-de\-pen\-dent $k-1$ first integrals and with the Jacobi multiplier $\mu$ may be rewritten in the Hamiltonian form
\begin{equation}\label{eqy-ham}
Y=P^{(y)}\mathrm d H\,,
\end{equation}
according to the Vall\'{e}e-Poussin theorem on functional determinants \cite{vp38}. Here $P^{(y)}$ is a rank-two Poisson bivector and $H$ is a function on $H_1,\ldots,H_{k-1}$, see details in \cite{bm99} .

For instance, we can put
\begin{equation}\label{y-poi}
P^{(y)}_{12}=\mu ^{-1}\,,\qquad P^{(y)}_{13}=Y_1\,,\qquad P^{(y)}_{23}=Y_2\,,
\end{equation}
so that
\[
\{y_1,y_2\}=\mu^{-1}\,,\qquad \dot{y}_1=\{y_1,H\}=Y_1\,,\qquad \dot{y}_2=\{y_2,H\}=Y_2\,,
\]
and
\[\dot{y_i}=0\,,\quad i\geq3\,.\]
Other brackets are equal to zero. The corresponding rank-two bivector $P^{(y)}$ is the Poisson bivector iff
\[
\dfrac{\partial \mu Y_1}{\partial y_1}+\dfrac{\partial \mu Y_2}{\partial y_2}=-{\mu^2Y_1^2}\,\dfrac{\partial }{\partial y_3} \dfrac{Y_2}{Y_1}\,.
\]
If we choose variables $y_1,\ldots,y_k$ so that the ratio $Y_2/Y_1$ is independent of $y_3$, then the Hamiltonization of dynamical equations (\ref{eqy-ham}) is equivalent to the existence of an invariant measure because
\[
[P^{(y)},P^{(y)}]=0\,,\qquad\Leftrightarrow\qquad \sum_{i=1}^k \dfrac{\partial \mu X_i}{\partial y_i}=0\,.
\]
Here $[.,.]$ is a Schouten bracket.

Such rank-two Poisson structures for many nonholonomic systems are well-studied \cite{fass05,herm95,mosh87,ram04}. Moreover, there are some examples of rank-four  Poisson structures for other nonholonomic systems \cite{bts12,bm01,ts12b}.

The main aim of this note is to discuss rank-two and rank-four Poisson structures for the nonholonomic St\"{u}bler model, which describes rolling  without sliding of a uniform ball on a cylindrical surface.

\section{Homogeneous ball on a surface}
\setcounter{equation}{0}
Let us consider a homogeneous ball with mass $m$, radius $R$ and tensor of inertia $\mathbf I = \mu\mathbf E$, where $\mathbf E$ is a unit matrix. We are going to study the case when the ball rolls without sliding on the surface having only one point in common with it during  the entire  motion.

In rigid body dynamics it is customary to introduce a body-fixed frame of reference. However, while
studying the motion of a homogeneous ball, it is more convenient to write the equations of motion with
respect to a certain frame of reference fixed in space \cite{bmk02}.

The surface is defined by the equation
\[\mathfrak f(r)=0\,\]
and the rolling ball is subject to two kinds of constraints: the holonomic constraint of motion on  the surface and no slip nonholonomic constraint associated with the zero velocity at the point of contact
\[v+\omega\times a=0\,,\]
where we denote the velocity of the ball's center by $v$, the angular velocity of the ball by $\omega$ and the vector joining the ball's center with the contact point by $a$.

If $N$ is the reaction force at the contact point, $F$ and $M_F$ are the external force and its moment with respect to the contact point, then the conservation principles of linear and angular momentum read as
\[
m\dot{v}=N+F\,,\qquad \dot{\mathbf I \omega}=a\times N+M_F\,.
\]
Substituting $a=-R\g$, where $\g$ is a normal to the surface, and eliminating the reaction force $N$, one gets the following equations of motion
\begin{equation}\label{ball-eqm}
\dot{M}=d\dot{\g}\times (\omega\times\g)+M_F\,,\qquad \dot{r}+R\dot{\g}=\omega\times R\g\,,
\end{equation}
which define vector field $X$ on the six-dimensional phase space with local coordinates $x=(\g_1,\g_2,\g_3,M_1,M_2,M_3)$. Here $d=mR^2$ and the vectors $\omega$ and $r$ are functions of $x$ defined by the following relations
\[
M=\mu\omega+d\g\times(\omega\times\g)\,,\qquad\g=\dfrac{\nabla \mathfrak f(r)}{|\nabla \mathfrak f(r)|}\,.
\]
Remind that all the vectors are expressed in the space frame \cite{bmk02}.

If external force $F$ is a potential force associated with potential $U=U(r+R\g)$, then
\[M_F=R\g\times\dfrac{\partial U}{\partial r'}\,,\qquad r'=r+R\g\,,
\]
and equations (\ref{ball-eqm}) possess first integrals
\[
H=\frac12(M,\omega)+U(r')\,,\qquad C=(\g,\g)=1\,.
\]
Below we consider  a homogeneous ball rolling on the inner side of the  cylindrical surface. Apparently Routh and St\"{u}bler were first to explore this problem  \cite{rou55,stub09} .

\subsection{St\"{u}bler model}
 Following  \cite{stub09} let us consider ball  on a cylindrical surface. As before,, the ball has only one poimt of the contact with the surface and, as usual,  it is assumed that there is  no slipping.

\begin{wrapfigure}[11]{r}{0.3\linewidth}
\hspace{6ex}\includegraphics[width=0.7\linewidth]{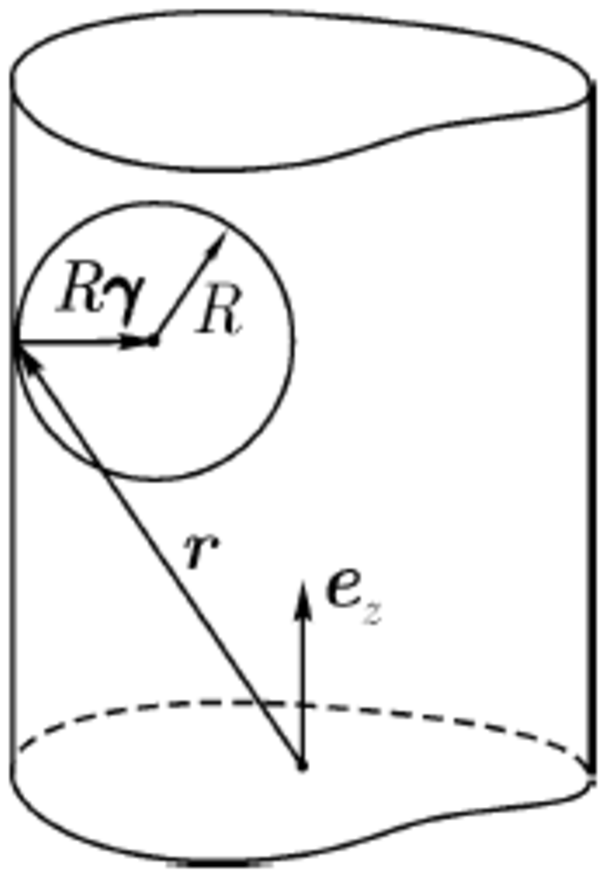}
\end{wrapfigure}
Let us choose a fixed frame of reference,  with the axis $Oz$ directed along axis forming the cylinder. In this case the symmetry group $G=SO (3) \times  T_z$ consists of  the ball's  rotations and of  parallel translations  along the $Oz$ axis.

We can add  potential  $U (z)$ depending only on the coordinate $z$ without loss of integrability of the system. So below  we will consider ball in a  gravitational field with potential
\[U(z)=-m\mathrm g z.\]
The vector of a normal to a surface at a point of contact looks like
\[
\g=(\g_1,\g_2,0)\,,\qquad \g_1^2+\g_2^2=1\,.
\]
The projections of the normal  $\tilde \g=(\g_1,\g_2)$ and of the  vector
of the center of mass of the ball   $\tilde r=(r_1+R\g_1,r_2+R\g_2)$ onto the normal cross-section satisfy to evident geometrical relations
\[(\dot{\tilde r},\tilde{\g})=(\dot{\tilde \g},\tilde{\g})=0\,.\]
Hence, we conclude that  $\dot{\tilde \g}$
 is parallel to $\dot{\tilde r}$, i.e.
\[
\dot{\tilde \g}=f(\g_1,\g_2)\dot{\tilde r}\,.
\]
The factor $f(\g_1,\g_2)\equiv f(\g)$ is completely determined by the geometry of the cylinder's cross-section and does not
depend on the angular velocity.

In this case equations  (\ref{ball-eqm}) look like
\ben\label{cyl-eqm}
\dot{M}&=&\dfrac{df(\g)M_3}{(d+\mu)^2}\,(M\times\g,e_z)\,\g-m\mathrm g\,e_z\times \g\,,\qquad d=m^2R\,,\\
\nn\\
\dot{\g}&=&\dfrac{f(\g)M_3}{d+\mu)}\,e_z\times\g\,,\qquad \dot{z}=\dfrac{1}{d+\mu}\,(M\times\g,e_z)\,.\nn
\en
Here  $e_z=(0,0,1)$ is a unit vector along the axis  $Oz$,  where $z=r_3/R$.

Equations of motion  (\ref{cyl-eqm}) define the six-dimensional vector field $X$ with the invariant volume form
\bq\label{mu-cyl}
\mu=f(\g)\mathrm d\g\mathrm dz\mathrm dM\,,
\eq
with three polynomial integrals of motion
\[H=\dfrac{1}{2}(M,\omega)+U(z)\,,\qquad C=(\g,\g)=1\,,\qquad F=M_3=(d+\mu)\omega_3\]
and two rational integrals \cite{stub09}
\bq\label{cyl-int}
J_{k}=v_1^{(k)}M_1+v_2^{(k)}M_2-\dfrac{m\mathrm g d}{(\nu^2-1)}\,\dfrac{ v_3^{(k)}}{M_3}\,,\qquad k=1,2\,.
\eq
Of course, multiplying these integrals by $M_3$, we obtain polynomial integrals of motion.

If
\[\nu=\sqrt{\dfrac{\mu}{d+\mu}}\,,\qquad \g_1=\cos\phi\,,\qquad \g_2=\sin\phi\,,\]
then the functions $v_{1,2}^{(k)}$ in  the two last integrals of motion
\bq\label{cyl-v}
\begin{array}{l}
v_1^{(k)}=\g_2^{3/2}L^{(k)}\left(\nu-\dfrac{1}{2},\dfrac{3}{2},\g_1\right)\,,\qquad
v_3^{(k)}=\displaystyle \int \dfrac{\g_1v_2^{(k)}-\g_2 v_1^{(k)}}{f(\phi)}\,\mathrm d \phi\,,\\
\\
v_2^{(k)}=\dfrac1{\g_2^{3/2}(\nu+1)}\left( L^{(k)}\left(\nu+\dfrac12,\dfrac32,\g_1\right)-\g_1\bigl(\g_2^2(\nu+1)+1\bigr)L^{(k)}\left(\nu-\dfrac12,\dfrac32,\g_1\right)\right)\\
\end{array}
\eq
are expressed by the associated Legendre functions of the first and second kinds $L^{(1,2)}$ and their integrals.

In fact any solution of (\ref{cyl-eqm}) may be obtained from the solution of a single equation
\bq\label{phi-eqm}
\dot{\phi}=ad^{-1}(1-\nu^2)  f(\phi)\,,
\eq
where $a$ is a value of the integral of motion $M_3$ and   $f(\phi)=f(\g)$ is a smooth bounded $2\pi$-periodic function.  Other variables are functions of $\phi$ and of the values of integrals of motion.

\section{Rank-two Poisson structures}
Let us pass from variables $x = (\g_1, \g_2, z, M_1, M_2, M_3) $ to variables $y = (y_1, \ldots, y_6) $ by the rule
\[
y_1=\g_1\,,\qquad y_2=\g_2\,,\qquad y_3=H\,,\qquad y_4=M_3\,,\qquad y_5=J_1\,,\qquad y_6=J_2\,.
\]
In the new variables the initial system of the equations (\ref {cyl-eqm}) has the form (\ref {eqy-in}) and the corresponding
functions
\[
Y_1=d^{-1}(\nu^2-1)f(y_1,y_2)\,y_2\,y_4\,,\qquad Y_2=-d^{-1}(\nu^2-1)f(y_1,y_2)\,y_1\,y_4
\]
are independent on the Hamiltonian $y_3=H$. So, we can introduce the rank-two Poisson bivector  $P^{(y)}$ (\ref{y-poi}), which allows us to identify initial vector field  $X$ (\ref{cyl-eqm})
\[X=P^{(y)}\mathrm d H\]
with the Hamiltonian vector field $P^{(y)}\mathrm d H$.

According to the construction of the Poisson bivector $P^{(y)}$,  its Casimir functions are equal to  $y_4=M_3$, $y_5= J_{1}$ and $y_6=J_2$
\[
P^{(y)}\mathrm d M_3=P^{(y)}\mathrm d J_1=P^{(y)} \mathrm d J_2=0\,,
\]
 whereas geometric integral of motion  $C=\g_1^2+\g_2^2=\cos^2\phi+\sin^2\phi$ is not a Casimir function and  by definition (\ref{y-poi}) we can prove that the Poisson bracket
\[\{\g_1,\g_2\}_y=f(\g_1,\g_2)\]
is equal to the density of the invariant measure.

So, we have some "non physical"\, rank-two Poisson structure, which allows us to rewrite initial vector field in the Hamiltonian form
\[X=P^{(y)}\mathrm d H\,.\]

\subsection{A circular cylinder}
Let us present  Poisson brackets associated with the  obtained Poisson bivector  $P^{(y)}$, in the special case
\[f(\g)=R_c\,.\]
 Here $R_c>R$ is the radius of the circular cylinder.  In this case we can easily express the initial $x$ variables via auxiliary (formal) variables $y$:
\ben
\g_1&=&y_1\,,\qquad \g_2=y_2\,,\qquad M_3=y_4\,,\nn\\
\nn\\
M_1&=&\frac{\Bigl(\nu y_1\cos\nu\arctan(y_2/y_1)+y_2\sin\nu\arctan(y_2/y_1)\Bigl)y_5}{\nu(y_1^2+y_2^2)}
+\frac{dm\mathrm g\, y_1}{(y_1^2+y_2^2)(\nu^2-1)R_cy_4}\nn\\
\nn\\
&+&\frac{\Bigl(\nu y_1\sin\nu\arctan(y_2/y_1)-y_2\cos\nu\arctan(y_2/y_1)\Bigr)y_6}{\nu(y_1^2+y_2^2)}
\,,\nn\\
M_2&=&\frac{\Bigl(\nu y_2\cos\nu\arctan(y_2/y_1)-y_1\sin\nu\arctan(y_2/y_1)\Bigl)y_5}{\nu(y_1^2+y_2^2)}
+\frac{dm\mathrm g\, y_2}{(y_1^2+y_2^2)(\nu^2-1)R_cy_4}\nn\\
\nn\\
&+&\frac{\Bigl(\nu y_2\sin\nu\arctan(y_2/y_1)+y_1\cos\nu\arctan(y_2/y_1)\Bigr)y_6}{\nu(y_1^2+y_2^2)}\nn\\
\nn\\
z&=&-\frac{1}{m\mathrm g}\left(y_3+\frac{(\nu^2-1)(\g_1+\nu^2\g_2^2)M_1^2}{2\nu^2d}-\frac{(\nu^2-1)^2\g_1\g_2M_1M_2}{\nu^2d}\right.\nn\\
\nn\\
&+&\left.\frac{(\nu^2-1)(\nu^2\g_1+\g_2^2)M_2^2}{2\nu^2d}+\dfrac{(\nu^2-1)y_4^2}{2d}\right)\,.\nn
\en
It allows us to get an explicit form of the Poisson brackets for the initial  $x$-variables. For instance, one gets
\ben
 \{\g_1,z\}_y&=&-\dfrac{(\nu^2-1)R_c\g_2M_3}{m\mathrm g d}+\dfrac{R_c(\nu^2-1)(\nu^2M_1^2+M_2^2)\g_2}{\nu^2m\mathrm g d}+\dfrac{\g_1(\g_2M_1-\g_1M_2)}{M_3}\,,\nn\\
 \nn\\
 \{\g_1,M_1\}_y&=&-\frac{(\nu^2\g_1^2+\g_2^2)R_c\g_2M_1}{(\g_1^2+\g_2^2)^2}
 +\frac{(\nu^2-1)R_c\g_1^3M_2}{(\g_1^2+\g_2^2)^2}
 -\frac{m\mathrm g d\g_1\g_2}{(\g_1^2+\g_2^2)^2(\nu^2-1)M_3}\,,\label{poi21}\\
 \nn\\
 \{\g_1,M_2\}_y&=&
 -\frac{(\nu^2-1)R_c\g_1\g_2^2M_1}{(\g_1^2+\g_2^2)^2}+\frac{(\nu^2\g_1^2-\g_2^2-2\g_1^2)R_c\g_2M_2}{(\g_1^2+\g_2^2)^2}
 +\frac{m\mathrm g d\g_1^2}{(\g_1^2+\g_2^2)^2(\nu^2-1)M_3}\,.\nn
\en
Other brackets are more bulky and we will omit here their explicit expressions for  brevity.

So, we obtain sufficiently complicated second-rank Poisson structure using the Euler-Jacobi theorem. Of course, for the St\"{u}bler problem with five integrals of motion we can get a few such Poisson structures.

\subsection{Second rank-two Poisson brackets}
A little more conventional from the kinematic point of view  Poisson brackets may be obtained in the following way.
Using reduction by the geometric integral of motion
\[C=\g_1^2+\g_2^2=\cos^2\phi+\sin^2\phi=1,\]
 we can  consider a vector field on the five-dimensional phase space with coordinates  $\hat{x}=(\phi,z,M_1,M_2,M_3)$ and only after that change variables  $\hat{x}\to y$:
\[
y_1=\phi\,,\qquad y_2=M_3\,,\qquad y_3=H\,,\qquad Y_4=J_1\,,\qquad y_5=J_2\,.
\]
In this case  equation  (\ref{phi-eqm}) for a circular cylinder looks like
\[Y_1=-\dfrac{R_c(\nu^2-1)y_2}{d}\,,\qquad Y_2=0\,.\]
In contrast with the previous case  the integral of motion  $M_3=y_2$ is not  a Casimir function. The corresponding Poisson brackets for the initial $x$ variables are:
\[
\{\phi,M_1\}_y=-\dfrac{m\mathrm g \cos\phi}{(\nu^2-1)M_3^2}\,,\qquad \{\phi,M_2\}_y=-\dfrac{m\mathrm g \sin\phi}{(\nu^2-1)M_3^2}\,,\qquad \{\phi,M_3\}_y=R_c\,,
\]
and
\ben
\{\phi,z\}_y&=&\dfrac{\cos\phi M_1+\sin\phi M_2}{\nu^2M_3^2}\,,\qquad \{M_3,z\}_y=-\dfrac{\sin\phi M_1+\cos\phi M_2}{M_3}\,,\nn\\
\{z,M_1\}_y&=&\dfrac{(\nu^2-1)\cos\phi\bigl((M_1^2-M_2^2)\sin2\phi-2M_1M_2\cos2\phi\bigr)}{2\nu^2M_3^2}\nn\\
&-&\dfrac{m\mathrm gd\bigl(M_1(\nu^2-1)\sin2\phi-(\nu^2\cos2\phi+\nu^2+1-\cos2\phi)M_2\bigr)}{2R_c\nu^2(\nu^2-1)M_3^3}\,,\nn\\
\nn\\
\{z,M_2\}_y&=&\dfrac{(\nu^2-1)\sin\phi\bigl((M_1^2-M_2^2)\sin2\phi-2M_1M_2\cos2\phi\bigr)}{2\nu^2M_3^2}\nn\\
&-&\dfrac{m\mathrm gd\bigl(M_2(\nu^2-1)\sin2\phi+(\nu^2\cos2\phi-\nu^2-1-\cos2\phi)M_1\bigr)}{2R_c\nu^2(\nu^2-1)M_3^3}\,,\nn\\
\{M_3,M_1\}_y&=&R_c(\nu^2-1)\cos\phi(\sin\phi M_1-\cos\phi M_2)+\dfrac{m\mathrm g d \sin\phi}{(\nu^2-1)M_3}\,,\nn\\
\{M_3,M_1\}_y&=-&R_c(\nu^2-1)\sin\phi(\sin\phi M_1-\cos\phi M_2)+\dfrac{m\mathrm g d \cos\phi}{(\nu^2-1)M_3}\,.\nn
\en
We have to emphasize  that one gets  Poisson brackets   rational in momenta depending on the potential, in contrast wits all the known Poisson brackets for other nonholonomic systems \cite{bts12,bm01,fass05,herm95,mosh87,ram04,ts12b}.

Nevertheless, this rank-two Poisson structure also allows us to identify the initial vector field with the Hamiltonian vector field.

\section{Rank-four Poisson structure}
\setcounter{equation}{0}
Let us come back from the  circular cylinder to the generic cylindrical surface.  In order to get a rank-four Poisson structure in  this paper we use the brute force approach proposed in \cite{ts11}.

 As above we start with the Euler-Jacobi theorem. Namely, six equations of motion  (\ref{cyl-eqm}) possess four integrals of motion
\[H_1=H\,,\qquad H_2=C\,,\qquad H_3=J_1\,,\qquad H_4=J_2\]
and the invariant measure (\ref{mu-cyl}) and, therefore, they are integrable by quadratures according to the Euler-Jacobi theorem.

If we  identify the common level surfaces of integrals $H_1\,,\ldots,H_4$ with the Lagrangian foliation of symplectic leaves of some unknown rank-four  Poisson bivector $P$, then  the following equations must hold:
 \begin{equation}\label{geom-eq}
[P,P]=0\,,\qquad \{H_k,H_m\}=\sum_{i,j=1}^6 P_{ij}\,\dfrac{\partial H_k}{\partial x_i}\dfrac{\partial H_m}{\partial x_j}=0\,,\qquad k,m=1,\ldots,4,
\end{equation}
where $[.,.]$ is the Schouten bracket. Remind, that the Schouten bracket $[A,B]$ of two bivectors $A$ and $B$
is a trivector whose entries in local coordinates $x$ on $\mathcal M$ are
\begin{equation}\label{sh-br}
[A,B]_{ijk}=-\sum\limits_{m=1}^{dim\, \mathcal M}\left(B_{mk}\dfrac{\partial A_{ij}}{\partial x_m}
+A_{mk} \dfrac{\partial B_{ij}}{\partial x_m}+\mathrm{cycle}(i,j,k)\right).
\end{equation}
If $[P,P]=0$, bivector $P$ is a Poisson bivector.

The brute force method consists of a direct solution of  equations (\ref{geom-eq}) with respect to entries of the bivector $P$ using an appropriate anzats. Recall that  a'priory these equations have infinitely many solutions \cite{ts07} and using an anzats we can obtain some partial solutions in a constructive way.   Below we use the following simple anzats
\begin{equation} \label{p-anz}
P_{ij}=\sum_{k=1}^6 u_{ij}^k(\g,z)M_k+w_{ij}(\g,z)\,,
\end{equation}
  where $c_{ij}^k$ and $d_{ij}$ are unknown functions of $\g$ and $z$.

In this class of bivectors (\ref{p-anz}) equations  (\ref{geom-eq}) have  nontrivial solutions  if and only if
\[U(z)=0.\]
In this case there are two nontrivial solutions
\bq\label{cyl-p}
P^{(k)}=\alpha(\g_1,\g_2,z)\left(
                             \begin{array}{cc}
                               0 & \mathbf \Gamma_\alpha \\
                               -\mathbf \Gamma_\alpha^\top & 0 \\
                             \end{array}
                           \right)+\beta(\g_1,\g_2)\left( \begin{array}{cc}
                               0 & \mathbf \Gamma_\beta \\
                               -\mathbf \Gamma_\beta^\top & \mathbf M_\beta \\
                             \end{array}
                           \right)\,,\quad k=1,2,
\eq
with the following Casimir functions
\[ P^{(k)}\mathrm d(\g,\g)=P^{(k)}\mathrm dJ_k=0\,,\qquad \mbox{rank}P^{(k)}=4\,.\]
 Here $\alpha$ and  $\beta$ are arbitrary functions,  matrices  $\mathbf \Gamma_\alpha$ and   $\mathbf \Gamma_\beta$ read as
\[
\mathbf \Gamma_\alpha=\left(
                        \begin{array}{ccc}
                          0 & 0 & 0 \\
                          0 & 0 & 0 \\
                          1 &\rho_k  & \sigma_k \\
                        \end{array}
                      \right)\,,\qquad
  \mathbf \Gamma_\beta=\left(
                        \begin{array}{ccc}
                          0 & 0 & -\g_2 \\
                          0 & 0 & \g_1 \\
                          0 &0 & 0 \\
                        \end{array}
                      \right)\]
and the skew-symmetric matrix  $\mathbf M_\beta$  depends only of  $M_1$ and $M_2$
\[
  \mathbf M_\beta=\left(
                        \begin{array}{ccc}
                          0 & 0 & \g_1(\g_1M_2-\g_2M_1)(\nu^2-1) \\
                          0 & 0 & \g_2(\g_1M_2-\g_2M_1)(\nu^2-1) \\
                          * &*& 0 \\
                        \end{array}
                      \right)\,.
\]
The functions appearing in the definition of  $\mathbf \Gamma_\alpha$ are 
\[
\rho_k=-\dfrac{v_1^{(k)}}{v_2^{(k)}}\,,\qquad \sigma_k=\int \left( \dfrac{\g_1(\nu^2-1)(\g_2-\g_1\rho_k)}{\alpha(\g_1,\g_2,z)}+\dfrac{\g_1\frac{\partial \alpha(\g_1,\g_2,z)}{\partial \g_2}-\g_2\frac{\partial\alpha(\g_1,\g_2,z)}{\partial \g_1}}{\alpha^2(\g_1,\g_2,z)} \right)\mathrm d z\,.
\]
Here $v_{1,2}^{(k)}$  (\ref{cyl-v})  are coefficients before  $M_{1,2}$ in the definition of the integrals of motion $J_{1,2}$ (\ref{cyl-int}).

\begin{prop}
For the homogeneous ball on a cylindrical surface the vector  field $X$ (\ref{cyl-eqm}) is a linear combination of the Hamiltonian vector field and the symmetry field $X_S$
\bq\label{cyl-dec}
X=f(\g)\beta^{-1}\,P^{(k)}\mathrm d H+\dfrac{\eta_k (\nu^2-1)}{d}\, X_S\,.\qquad
\eq
The  symmetry field
 \[
 X_S=\,\dfrac{\partial}{\partial z}\,,
 \]
 describes the parallel translations  along  the axis $Oz$ and the coefficients in  (\ref{cyl-dec}) are
 \ben
 \eta_k&=&(\g_1M_2-\g_2M_1)\nn\\
&+&\alpha f(\g)\left(
 \sigma_k M_3+(\g_1\rho_k-\g_2)(\g_1M_2-\g_2M_1)+\frac{(\g_2\rho_k+\g_1)(\g_1M_1+\g_2M_2)}{\nu^2}
 \right)\,.\nn
 \en
\end{prop}
For a few other nonholonomic systems  similar decompositions of the initial  vectors field on the Hamiltonian and symmetry vector field is discussed in the review \cite{bmt13}.

\section{Conclusion}
We discuss two constructions of the Poisson bivectors based on the Euler-Jacobi theorem  for the nonholonomic St\"{u}bler model.  Rank-two Poisson bivectors may be considered as some mathematical toys in the theory of deformations of  the canonical Poisson brackets.

Rank-four Poisson bivectors are more interesting. Remind that if we identify the  common level surfaces of integrals of motion with Lagrangian foliation with respect to some Poisson bivector later we can try to  identify the  common level surfaces of integrals of motion with bi-Lagrangian foliation with respect to some Poisson pencil and then with some subvariety of the Jacoby variety associated with some algebraic curve related with this pencil. Of course, in this concrete model we can directly solve equations of motion without of these sophisticated mathematical tools.

This work was partially supported by RFBR grant 13-01-00061.

\end{document}